\documentclass[prb,twocolumn]{revtex4-1}

\usepackage[pdftex]{graphics}
\usepackage{amssymb,amsmath}

\begin{document}

\title{Nuclear quantum effects in thermal conductivity from centroid molecular dynamics}
\author{Benjamin J. Sutherland}
\affiliation{Department of Chemistry, University of Oxford, Physical and Theoretical Chemistry Laboratory, South Parks Road, Oxford, OX1 3QZ, UK}
\author{William H. D. Moore}
\affiliation{Department of Chemistry, University of Oxford, Physical and Theoretical Chemistry Laboratory, South Parks Road, Oxford, OX1 3QZ, UK}
\author{David E. Manolopoulos}
\email{david.manolopoulos@chem.ox.ac.uk}
\affiliation{Department of Chemistry, University of Oxford, Physical and Theoretical Chemistry Laboratory, South Parks Road, Oxford, OX1 3QZ, UK}
\begin{abstract}
We show that the centroid molecular dynamics (CMD) method provides a realistic way to calculate the thermal diffusivity $a=\lambda/\rho\,c_{\rm V}$ of a quantum mechanical liquid such as para-hydrogen. Once $a$ has been calculated, the thermal conductivity can be obtained from $\lambda=\rho\,c_{\rm V}a$, where $\rho$ is the density of the liquid and $c_{\rm V}$ is the constant-volume heat capacity. The use of this formula requires an accurate quantum mechanical heat capacity $c_{\rm V}$, which can be obtained from a path integral molecular dynamics simulation. The thermal diffusivity can be calculated either from the decay of the equilibrium density fluctuations in the liquid or by using the Green-Kubo relation to calculate the CMD approximation to $\lambda$ and then dividing this by the corresponding approximation to $\rho\, c_{\rm V}$. We show that both approaches give the same results for liquid para-hydrogen and that these results are in good agreement with experimental measurements of the thermal conductivity over a wide temperature range. In particular, they correctly predict a decrease in the thermal conductivity at low temperatures -- an effect that stems from the decrease in the quantum mechanical heat capacity and has eluded previous para-hydrogen simulations. We also show that the method gives equally good agreement with experimental measurements for the thermal conductivity of normal liquid helium.
\end{abstract}

\maketitle

\section{Introduction}

Allowing for nuclear quantum effects in calculations of thermal conductivity has proved to be a challenging problem in molecular simulation. For crystalline solids, one can use anharmonic lattice dynamics simulations in which the anharmonicities in the interaction potential that lead to phonon scattering are treated perturbatively and nuclear quantum effects are incorporated via the harmonic quantum mechanical phonon distribution function.\cite{Ladd86,Turney09} However, for strongly anharmonic and disordered systems such as liquids and glasses, this approach is inapplicable, and we are not aware of any existing method that is able to reproduce experimental measurements of the thermal conductivity when nuclear quantum effects play a significant role.

A number of approximate methods have been developed for the calculation of (Kubo-transformed\cite{Kubo57}) quantum mechanical correlation functions, including centroid molecular dynamics (CMD),\cite{Cao94a,Cao94b,Cao94c,Cao94d,Cao94e,Jang99a,Jang99b} ring polymer molecular dynamics (RPMD),\cite{Craig04} and the linearised semiclassical initial value representation (LSC-IVR).\cite{Miller01} All three of these have been used\cite{Yonetani04,Liu11,Luo20} to calculate thermal conductivities from the Green-Kubo relation
\begin{equation}
\lambda = {1\over 3Vk_{\rm B}T^2}\int_0^{\infty}\left[{1\over\beta\hbar}\int_0^{\beta\hbar}\left<\hat{\bf J}(-i\tau)\cdot\hat{\bf J}(t)\right> {\rm d}\tau\right]{\rm d}t,
\end{equation}
in which $\hat{\bf J}$ is the energy current operator, $\hat{\bf J}(t) = e^{+i\hat{H}t/\hbar}\hat{\bf J}e^{-i\hat{H}t/\hbar}$, and the angular brackets denote a canonical quantum mechanical average
\begin{equation}
\left<\cdots\right> = {\rm tr}\left[e^{-\beta\hat{H}}\left(\cdots\right)\right]/{\rm tr}\left[e^{-\beta\hat{H}}\right]
\end{equation}
with $\beta=1/k_{\rm B}T$. However, none of the above methods is expected to provide an especially reliable way to calculate the correlation function in Eq.~(1). 

The CMD and RPMD approximations are expected to be most reliable when the operators in the Kubo-transformed correlation function are linear functions of the atomic position and momentum operators. This is not the case for the energy current operators in Eq.~(1), which can be written as
\begin{equation}
\hat{\bf J} = {1\over 2}\sum_{i=1}^N {i\over\hbar}\left[\hat{H},\hat{E}_i\,\hat{\bf r}_i+\hat{\bf r}_i\,\hat{E}_i\right],
\end{equation}
where for a system with pairwise interactions
\begin{equation}
\hat{E}_i = {|\hat{\bf p}_i|^2\over 2m_i}+{1\over 2}\sum_{j\not= i}^N v(|\hat{\bf r}_i-\hat{\bf r}_j|).
\end{equation}  
Indeed the structure of the operator $\hat{\bf J}$ is such that neither the CMD approximation\cite{Yonetani04} nor the RPMD approximation\cite{Luo20} even gives the correct Kubo-transformed energy current autocorrelation function at time zero.

The LSC-IVR approximation does (in principle) give the correct correlation function at time zero, by treating the Wigner transform of the initial Kubo-transformed energy current operator
\begin{equation}
\hat{\bf J}_{\beta} = {1\over \beta\hbar}\int_0^{\beta\hbar} \hat{\bf J}(-i\tau)\,{\rm d}\tau
\end{equation}
correctly.\cite{Liu11} However, the practical evaluation of the Wigner transform of $\hat{\bf J}_{\beta}$ requires additional approximations,\cite{Liu11} and more importantly, the subsequent dynamics in the LSC-IVR is inconsistent with the Wigner-transformed initial condition, because classical molecular dynamics does not conserve the Wigner transform of the Boltzmann operator. This leads to unphysical effects in the simulation that can become quite pronounced over the long time scales that are needed to calculate transport coefficients.\cite{Habershon09}

\begin{figure}[t]
\centering
\resizebox{0.8\columnwidth}{!} {\includegraphics{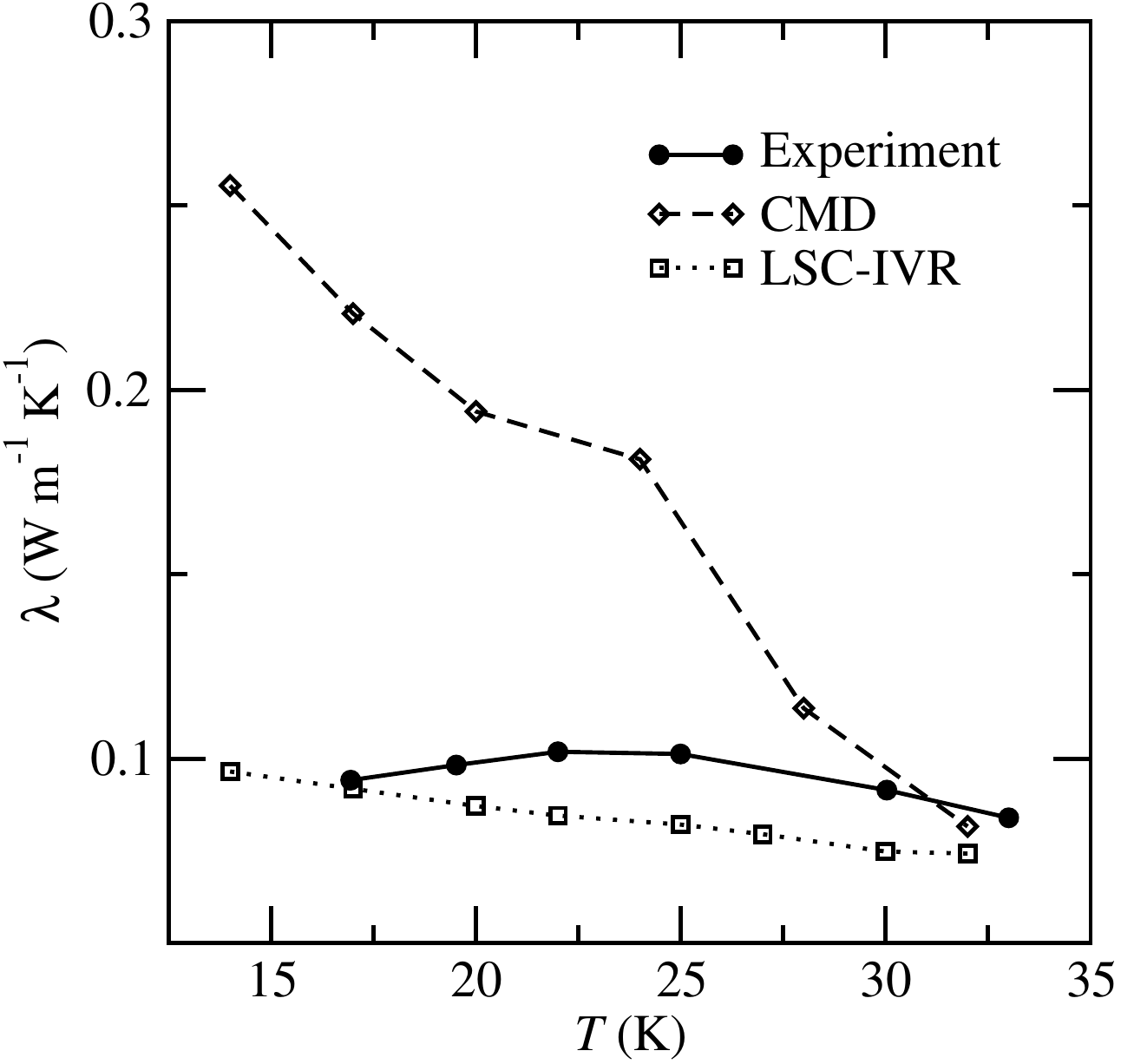}}
\caption{Comparison of CMD (Ref.~\onlinecite{Yonetani04}), LSC-IVR (Ref.~\onlinecite{Liu11}), and experimental (Ref.~\onlinecite{Roder70}) thermal conductivities of liquid para-hydrogen at densities close to the saturation line.}
 \end{figure}

To illustrate these shortcomings, we show in Fig.~1 the results of two previous simulations of the thermal conductivity of liquid para-hyrdogen. The CMD results of Yonetani and Kinugawa\cite{Yonetani04} and the LSC-IVR results of Liu {\em et al.}\cite{Liu11} were both calculated using the Green-Kubo relation, for liquids close to the saturation line. The experimental results of Roder and Diller\cite{Roder70} were measured under similar conditions. One sees that both calculations give thermal conductivities that increase monotonically with decreasing temperature, whereas the experimental thermal conductivity has a maximum at around 22 K and decreases at lower temperatures. A similar decrease in the thermal conductivity at low temperatures is also seen in other systems, including both liquids\cite{Donnelly98} and solids.\cite{Glassbrenner64} The decrease is a clear indication of the participation of nuclear quantum effects, and our goal here is to develop a method that is capable of capturing it.

In order to do this, we shall stick with para-hydrogen as our main example. This is an ideal model system with which to illustrate the role of nuclear quantum effects, for a variety of reasons. The thermal equilibrium population of the first excited $J=2$ rotational state of para-hydrogen is less than one part in a million at the critical point ($T_{\rm c}\approx 33.1$ K), so the para-hydrogen molecules are in their ground rotational state throughout the liquid state. Since the $J=0$ rotational wave function is spherically symmetric, the interaction between para-hydrogen molecules can be modelled to a good approximation by a scalar pair potential. The de Broglie thermal wavelength of the para-hydrogen molecules is sufficiently small at the triple point ($T_{\rm t}\approx 13.8$ K) that identical particle exchange effects can be ignored, and yet sufficiently large that it is essential to include quantum mechanical effects in the nuclear motion in the simulations.\cite{Miller05a} And finally, there is a wealth of high-quality experimental data available with which to check the accuracy of these simulations.\cite{Roder70,Eselson71,Roder73}  

We begin in Sec.~II by describing a fast implementation of CMD for liquid para-hydrogen and showing that it gives a realistic description of the diffusion of the molecules in the liquid, in agreement with previous studies.\cite{Yonetani04,Hone04,Hone05} The study of Yonetani and Kinugawa\cite{Yonetani04} also found that CMD provides a reasonable approximation to the experimental shear and bulk viscosities, although not to the thermal conductivity (see Fig.~1). The reason why CMD works better for diffusion than for thermal conductivity is that the velocity autocorrelation function that enters the Green-Kubo relation for the diffusion coefficient involves linear momentum operators. The CMD approximation is better justified for this linear correlation function than it is for the non-linear energy current correlation function that is used to calculate the thermal conductivity.

Motivated by this observation, we proceed in Sec.~III to use CMD to calculate the thermal diffusivity from the relaxation of equilibrium density fluctuations. This involves calculating the (Kubo-transformed) intermediate scattering function $\tilde{F}(k,t)$ in the hydrodynamic ($k\to 0$) limit and fitting it to an appropriate functional form.\cite{Mountain66} Since the intermediate scattering function involves the correlation of the density operators $\sum_{i=1}^N e^{\pm i {\bf k}\cdot\hat{\bf r}_i}$, which become approximately linear functions of the position operators $\hat{\bf r}_i$ in the hydrodynamic limit, one might hope that this approach would provide a more reliable way to extract the thermal conductivity from CMD.  We find that, while the CMD approximation to $\tilde{F}(k,t)$ is not perfect, it does indeed provide a reasonable estimate of $a=\lambda/\rho\, c_{\rm V}$, in that when this is combined with an exact path integral molecular dynamics (PIMD) calculation of $c_{\rm V}$ it gives a thermal conductivity $\lambda=\rho\, c_{\rm V}a$ that is in much better agreement with experiment than the Green-Kubo CMD result in Fig.~1.

The hydrodynamic calculation turns out to be quite expensive, because of the need to integrate for long times to obtain a hydrodynamic fit to $\tilde{F}(k,t)$ and the need to extrapolate to large system sizes to reach the limit as $k\to 0$. We therefore return to the Green-Kubo expression for $\lambda$ in Sec.~IV, and show how to extract the CMD approximation to $a$ from that instead. This can be done simply by calculating the constant-volume heat capacity $c_{\rm V}$ of the classical CMD system along with the Green-Kubo CMD approximation to $\lambda$, and defining $a({\rm CMD)} = \lambda({\rm CMD)}/\rho c_{\rm V}({\rm CMD})$. The resulting $a({\rm CMD})$ is found to agree with that obtained from the hydrodynamic calculation to within the statistical error bars of the two simulations, so when this is used to calculate the thermal conductivity as $\lambda({\rm Scaled}) = \rho\, c_{\rm V}({\rm PIMD})a({\rm CMD)}\equiv \bigl[c_{\rm V}({\rm PIMD)}/c_{\rm V}({\rm CMD})\bigr]\lambda({\rm CMD)}$, the result again agrees well with experiment.

We end Sec.~IV by bringing all of this together and using the Green-Kubo method with $c_{\rm V}({\rm PIMD)}/c_{\rm V}({\rm CMD)}$ scaling to calculate the thermal conductivity of liquid para-hydrogen throughout the temperature range shown in Fig.~1. The results are found to agree well with the experimental measurements of Roder and Diller\cite{Roder70} at their thermodynamic state points. Finally, to show that this is not just a coincidence, we report a similar calculation for normal liquid helium between 3 and 4 K for which the agreement with experiment is equally compelling. Sec.~V contains some concluding remarks.

\section{Centroid molecular dynamics of liquid para-hydrogen}

\subsection{Theory}

At an operational level, the CMD approximation is simply classical molecular dynamics on an effective potential: the potential of mean force experienced by the centroid of the ring polymer in an imaginary time path integral simulation. This is a more modern alternative to earlier effective classical potentials that goes beyond the free ring polymer assumption of the Feyman-Hibbs approximation\cite{Feynman65} and avoids the locally harmonic assumption of the Feynman-Kleinert approximation.\cite{Feynman86} The path integral calculation that generates the centroid potential of mean force is done at the same thermodynamic state point as the CMD calculation, so the effective classical potential in CMD depends on both the temperature and the density of the system under investigation. 

The centroid potential of mean force is usually calculated \lq\lq on the fly" during a PIMD simulation, using the adiabatic CMD algorithm.\cite{Cao94d} However, this is computationally expensive, because it requires the use of a small time step to correctly integrate the rapid vibrations of the internal modes of the ring polymer when they are adiabatically separated from the centroid. To avoid this expense, Hone {\em et al.} have developed a \lq\lq fast CMD" method,\cite{Hone05} in which the forces obtained during a short PIMD simulation are least-squares fit to a pairwise model for the deviation between the classical interaction potential and the centroid potential of mean force. Here we shall use a similar approach that is tailored to the specific case of a system like liquid para-hydrogen, in which the classical interaction potential is a sum of pairwise contributions. This allows us to construct the pairwise approximation to the centroid potential of mean force without any force matching or least-squares fitting, as we shall now describe.

Suppose that $v(r)$ is the classical pair potential between two para-hydrogen molecules. Then the force between the centroids of the ring polymers of molecules $i$ and $j$ at a given configuration in a $P$-bead imaginary time path integral simulation is
\begin{equation}
{\bf f}_{ij} ^{(c)} = - {1\over P}\sum_{p=1}^P {{\bf r}_{ij}^{(p)}\over r_{ij}^{(p)}} v'(r_{ij}^{(p)}),
\end{equation}
where ${\bf r}_{ij}^{(p)}={\bf r}_i^{(p)}-{\bf r}_j^{(p)}$ is the vector from the $p$-th bead of molecule $j$ to the $p$-th bead of molecule $i$, and $r_{ij}^{(p)}=|{\bf r}_{ij}^{(p)}|$. The radial component of this force along the centroid-to-centroid vector
\begin{equation}
{\bf r}_{ij}^{(c)} = {1\over P}\sum_{p=1}^P {\bf r}_{ij}^{(p)}
\end{equation} 
is
\begin{equation}
f^{(c)}(r_{ij}^{(c)}) = - {1\over P}\sum_{p=1}^P {{\bf r}_{ij}^{(c)}\cdot{\bf r}_{ij}^{(p)}\over r_{ij}^{(c)}r_{ij}^{(p)}} v'(r_{ij}^{(p)}),
\end{equation}
and the average of this force over the configurations visited in the path integral simulation gives the pairwise approximation to the centroid potential of mean force from an integration:
\begin{equation}
v^{(c)}(r) = \int_r^{\infty} \left<f^{(c)}(r_{ij}^{(c)})\right>\,{\rm d}r_{ij}^{(c)}.
\end{equation}

Note that this calculation is significantly simplified by the fact that the interactions between all pairs of molecules are the same. If the system contains $N$ molecules, each snapshot of the PIMD simulation provides $O(N^2)$ estimates of $f^{(c)}(r_{ij}^{(c)})$ at different centroid-to-centroid distances $r_{ij}^{(c)}$, all of which contribute to the accumulation of the averages $\bigl<f^{(c)}(r_{ij}^{(c)})\bigr>$ in Eq.~(9). The statistical errors in the computed $v^{(c)}(r)$ are therefore small even after a relatively short simulation, and since one can use the same cutoff radius for $v^{(c)}(r)$ as for $v(r)$ [i.e., one can replace $\infty$ with $r_{\rm cut}$ in Eq.~(9)], there is no need to use an especially large system size in the PIMD calculation.

In practice, one can accumulate $\bigl<f^{(c)}(r_{ij}^{(c)})\bigr>$ within each histogram bin at the same time as calculating the PIMD centroid radial distribution function, and then do the integration in Eq.~(9) using the midpoint rule. This gives $v^{(c)}(r)$ on an equally-spaced grid of $r$ values that can be fit to a cubic spline for use in the subsequent CMD calculation. Since this CMD calculation is simply classical molecular dynamics with the effective Hamiltonian
\begin{equation}
H = \sum_{i=1}^N {|{\bf p}_i|^2\over 2m}+{1\over 2}\sum_{i=1}^N\sum_{j\ne i}^N v^{(c)}(r_{ij}),
\end{equation}
it too is very simple, and since this effective Hamiltonian only contains pairwise interactions there is no difficulty in using it to calculate a thermal conductivity from the Green-Kubo relation.

\subsection{Results and Discussion}

In order to validate this implementation of the fast CMD method, we have used it to perform some preliminary calculations at the two thermodynamic state points considered by Hone {\em et al.},\cite{Hone04} which have also been used in previous RPMD simulations.\cite{Miller05b} The first of these state points is a dense liquid near the triple point at $T=14$ K and $V=25.6$ cm$^3$ mol$^{-1}$, and the second is a sub-critical liquid at $T=25$ K and $V=31.7$ cm$^3$ mol$^{-1}$. Our PIMD calculations were performed with $P=64$ replicas of a system of $N=256$ para-hydrogen molecules, using a strongly stable Cayley integrator\cite{Korol19} combined\cite{Korol20} with a path integral Langevin equation thermostat,\cite{Ceriotti10} a time step of 2 fs, and a simulation time of 100 ps. The subsequent CMD calculations were performed with the same system size and time step. We used the isotropic part of the Silvera-Goldman\cite{Silvera78} pair potential for $v(r)$, truncated and shifted at $r_{\rm cut}=15$ bohr.

\begin{figure}[t]
\centering
\resizebox{0.8\columnwidth}{!} {\includegraphics{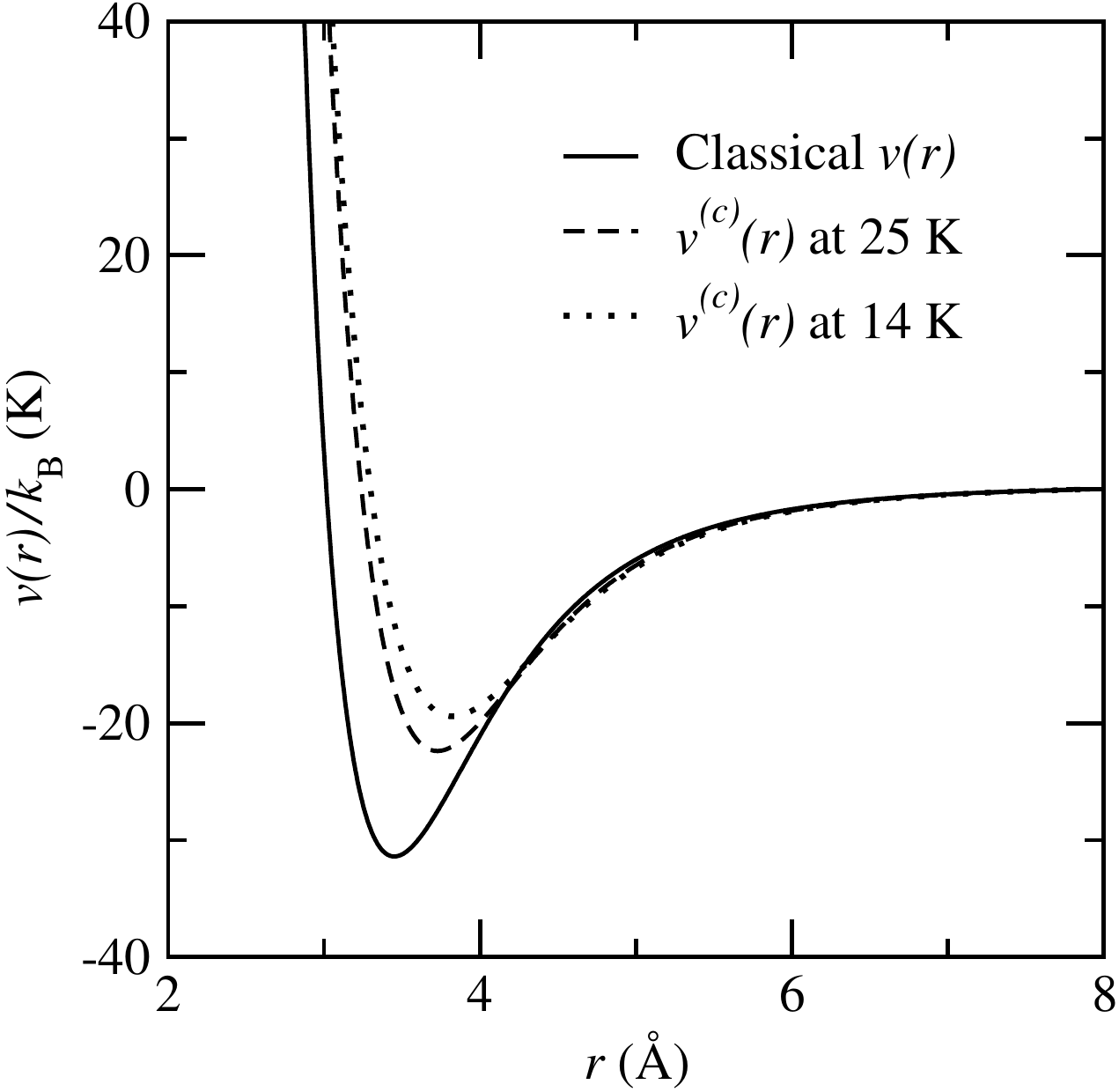}}
\caption{Comparison of the classical pair potential $v(r)$ of liquid para-hydrogen with the pairwise centroid potentials of mean force $v^{(c)}(r)$ at the $T=25$ K, $V=31.7$ cm$^{3}$ mol$^{-1}$ and $T=14$ K, $V=25.6$ cm$^{3}$ mol$^{-1}$ state points.}
 \end{figure}

Fig.~2 compares this classical pair potential with the pairwise centroid potentials of mean force obtained from Eq.~(9) at the two thermodynamic state points. These potentials of mean force are very similar to those in Fig.~1 of the paper by Hone {\em et al.},\cite{Hone05} who used their more general version of the fast CMD method involving least-squares force-matching. The nuclear quantum effects encapsulated in $v^{(c)}(r)$ are seen to increase the repulsion between the para-hydrogen molecules at short range and to increase the attraction between the molecules at long range, both effects being more pronounced at the more \lq\lq quantum mechanical" state point (14 K). These results are to be expected from the well-known \lq\lq swelling" of ring polymers at low temperatures due to thermal quantum fluctuations.\cite{Chandler81}

\begin{figure}[t]
\centering
\resizebox{0.8\columnwidth}{!} {\includegraphics{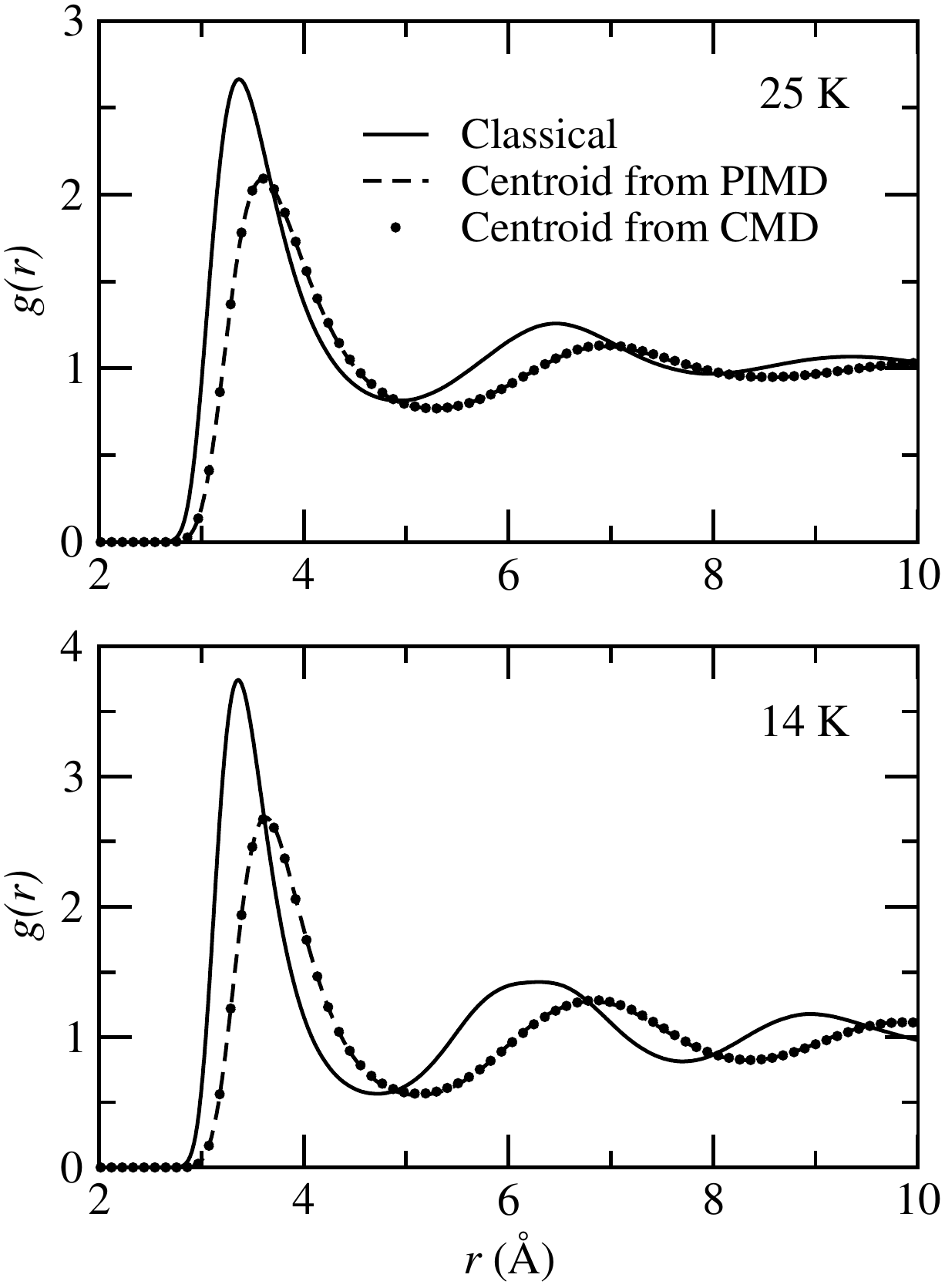}}
\caption{Classical and centroid (PIMD and CMD) radial distribution functions for liquid para-hydrogen at the $T=25$ K, $V=31.7$ cm$^{3}$ mol$^{-1}$ and $T=14$ K, $V=25.6$ cm$^{3}$ mol$^{-1}$ state points.}
 \end{figure}

Fig.~3 compares the centroid radial distribution functions at the two state points obtained from the PIMD and CMD simulations with those obtained from a purely classical simulation with the Silvera-Goldman pair potential. Although they could have been obtained by histogram binning as mentioned above, all of these radial distribution functions were actually computed using a recently-developed low-variance force estimator,\cite{Borgis13} which gives significantly better converged results for the same simulation time.\cite{Rotenberg20} As was noted by Hone {\em et al.},\cite{Hone05} the essentially perfect agreement between the centroid radial distribution functions obtained from the PIMD and CMD simulations confirms the validity of using a pairwise approximation to the centroid potential of mean force. We would add here that the difference between these centroid radial distribution functions and the purely classical radial distribution functions implies that liquid para-hydrogen cannot be treated classically at either thermodynamic state point. While both state points are in the liquid region of the quantum mechanical phase diagram, close to the liquid-vapour coexistence line, they are in the liquid-vapour coexistence region of the classical phase diagram.\cite{Miller05a} 

\begin{figure}[t]
\centering
\resizebox{0.8\columnwidth}{!} {\includegraphics{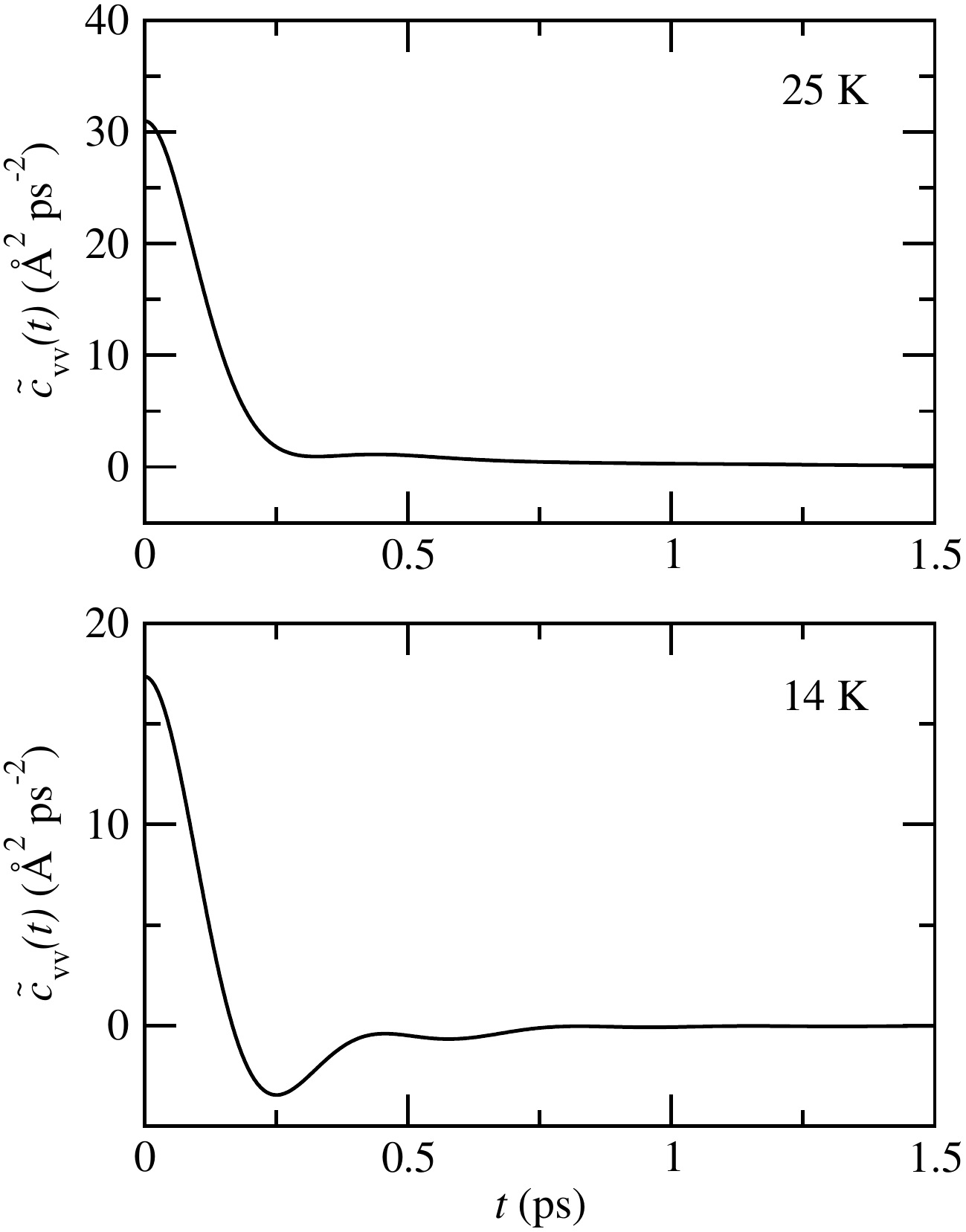}}
\caption{CMD velocity autocorrelation functions of liquid para-hydrogen at the $T=25$ K, $V=31.7$ cm$^{3}$ mol$^{-1}$ and $T=14$ K, $V=25.6$ cm$^{3}$ mol$^{-1}$ state points, as obtained from the present fast CMD simulations of a system of 256 para-hydrogen molecules.}
 \end{figure}

Finally, to check that the dynamics on the pairwise centroid potentials of mean force are also reasonable, we have calculated the CMD velocity autocorrelation functions and diffusion coefficients at both state points. These calculations were performed by running 1,000 separate 2 ps classical trajectories on each centroid potential of mean force, with a resampling of the momenta from the Maxwell distribution between each one. The resulting velocity autocorrelation functions are shown in Fig.~4. They are very similar to those obtained by Yonetani and Kinugawa using the adiabatic CMD algorithm,\cite{Yonetani04} and to those obtained by Miller and Manolopoulos using RPMD.\cite{Miller05b} 

The diffusion coefficients 
\begin{equation}
D={1\over 3}\int_0^{\infty} \tilde{c}_{{\bf v}\cdot{\bf v}}(t)\,{\rm d}t
\end{equation}
obtained from our velocity autocorrelation functions are $D=0.30$ \AA$^2$/ps at the 14 K state point and $D=1.49$ \AA$^2$/ps at 25 K. These results are in good agreement with those of Hone {\em et al.},\cite{Hone05} who applied both the adiabatic CMD algorithm and their force-matched (FM) implementation of fast CMD to a system of similar size to the one we have considered here (with 216 para-hydrogen molecules rather than our 256). Both our and their diffusion coefficients are slightly below the experimental\cite{Eselson71} values of 0.4  \AA$^2$/ps at 14 K and 1.6 \AA$^2$/ps at 25 K. In part this is because we have not corrected our results for finite-size effects, which are known to lead to an underestimation of $D$ in simulations with small system sizes.\cite{Dunweg93,Yeh04} Yonetani and Kinugawa's adiabatic CMD diffusion coefficients  are also slightly below the experimental values,\cite{Yonetani04} for the same reason. 

In addition to comparing diffusion coefficients with experiment, there is another way to check the accuracy of the velocity autocorrelation functions in Fig.~4.\cite{Miller05b} The CMD method provides an approximation to the Kubo-transformed velocity autocorrelation function
\begin{equation}
\tilde{c}_{{\bf v}\cdot{\bf v}}(t) = {1\over N}\sum_{i=1}^N {1\over \beta\hbar}\int_0^{\beta\hbar} \left<\hat{\bf v}_i(-i\tau)\cdot \hat{\bf v}_i(t)\right>{\rm d}\tau,
\end{equation}
in which $\hat{\bf v}_i=\hat{\bf p}_i/m_i$ is the velocity operator of molecule $i$ in the liquid and the angular brackets denote a canonical average as in Eq.~(2). The standard quantum mechanical velocity autocorrelation function
\begin{equation}
c_{{\bf v}\cdot{\bf v}}(t) = {1\over N}\sum_{i=1}^N \left<\hat{\bf v}_i(0)\cdot \hat{\bf v}_i(t)\right>
\end{equation}
is proportional to the average kinetic energy per molecule at time zero, $c_{{\bf v}\cdot{\bf v}}(0) = (2/m)\left<{\rm KE}\right>$, and this can be calculated from $\tilde{c}_{{\bf v}\cdot{\bf v}}(t)$ as\cite{Braams06}
\begin{equation}
c_{{\bf v}\cdot{\bf v}}(0) = \tilde{c}_{{\bf v}\cdot{\bf v}}(0)+\int_0^{\infty} {2\over (1-e^{+2\pi t/\beta\hbar})}{{\rm d}\tilde{c}_{{\bf v}\cdot{\bf v}}(t)\over {\rm d}t}\,{\rm d}t.
\end{equation}
Hence we can obtain an estimate of the average kinetic energy per molecule from the CMD approximation to $\tilde{c}_{{\bf v}\cdot{\bf v}}(t)$, and compare this with the exact quantum mechanical kinetic energy per molecule obtained by averaging the centroid virial kinetic energy estimator\cite{Herman82} over the configurations visited in a PIMD simulation. 

The results of this comparison are summarised in Table~I, along with the earlier results obtained by Miller and Manolopoulos using RPMD.\cite{Miller05b} One sees that, while the RPMD approximation to $\tilde{c}_{{\bf v}\cdot{\bf v}}(t)$ overestimates the kinetic energy, the CMD approximation underestimates it. However, the overestimation and underestimation are only slight, and both are far smaller than the errors in the purely classical kinetic energies $\left<{\rm KE}\right>_{\rm cl} = 3k_{\rm B}T/2$. Both approximations are therefore reasonable for liquid para-hydrogen according to this test of the accuracy of their Kubo-transformed velocity autocorrelation functions. We have chosen to use CMD for the present study because we shall need to use large system sizes in some of our calculations, and fast CMD is much cheaper than RPMD.

\begin{table}[b]
\begin{center}
\caption{Average kinetic energies per molecule
of liquid para-hydrogen at the $T=25$ K, $V=31.7$ cm$^3$ mol$^{-1}$
and $T=14$ K, $V=25.6$ cm$^3$ mol$^{-1}$ state points, as obtained by various methods 
for a system of 256 para-hydorgen molecules.}
\bigskip
\begin{tabular}{ccccc} \hline
&&& $\left<{\rm KE}\right>/k_{\rm B}$ in K\\
\  $T$ \ &\ \ \ CMD$^{(a)}$ \ \ \ & PIMD$^{(a)}$ & RPMD$^{(b)}$ & Classical \\ \hline
25 K  &61.6 & 62.3 & 64.2 & 37.5 \\
14 K &60.6 & 63.8 & 67.6 & 21 \\
\hline
\end{tabular}
\end{center}
$(a)$ Present calculations with $P=64$.\\
$(b)$ $P=24$ at 25 K and $P=48$ at 14 K, from Ref.~\onlinecite{Miller05b}.
\end{table}

Overall, we feel that the present results, and those of the earlier CMD studies we have mentioned,\cite{Yonetani04,Hone04,Hone05} establish that the CMD approximation is perfectly adequate for describing the diffusion of the molecules in liquid para-hydrogen. Especially when one considers that purely classical molecular dynamics simulations at the two state points we have considered bear so little relation to reality.\cite{Miller05a} The present calculations have also shown that our implementation of fast CMD in Eqs.~(8) and~(9) works correctly for a system with pairwise interactions. We shall therefore now move on to the more challenging problem of using this implementation to calculate the thermal conductivity of liquid para-hydrogen, first from the relaxation of equilibrium density fluctuations in Sec.~III and then from the Green-Kubo relation in Sec.~IV. 
 
\section{Thermal diffusivity from equilibrium density fluctuations}

\subsection{Theory}

It has been known for many years that the thermal diffusivity of a liquid is one of the key parameters that governs its equilibrium density fluctuations. The seminal study in this area was that of Mountain,\cite{Mountain66} who used an earlier suggestion of Landau and Placzek\cite{Landau34} to develop a theory for the time-dependence of the density fluctuations from the linearized hydrodynamic equations of irreversible thermodynamics. The resulting theory for the behaviour of the intermediate scattering function $F(k,t)$ in the hydrodynamic ($k\to 0$) limit has since been used extensively in the analysis of light scattering data,\cite{Berne70,Berne76} and discussed in textbooks.\cite{Hansen86,Boon91} It has also begun to be considered as computational tool -- as a way to actually extract the thermal diffusivity and other hydrodynamic parameters from simulations of the intermediate scattering function. The first study we are aware of along these lines was that of Schoen {\em et al.},\cite{Schoen86} who demonstrated that the hydrodynamic expression for $F(k,t)$ provides a viable way to calculate the thermal diffusivity and the sound attenuation coefficient of a Lennard-Jones liquid. Cheng and Frenkel have recently made some developments to the approach, applied it to a wider variety of systems, and advertised it as a practical way to calculate thermal diffusivities and thermal conductivities in situations where the Green-Kubo method is inapplicable.\cite{Cheng20}

Since we are interested here in the role of nuclear quantum effects, we shall interpret Mountain's formula for $F(k,t)$ as applying to the hydrodynamic limit of the Kubo-transformed intermediate scattering function
\begin{equation}
\tilde{F}(k,t) = {1\over N}{1\over \beta\hbar}\int_0^{\beta\hbar} \left<\hat{\rho}_{-{\bf k}}(-i\tau)\hat{\rho}_{\bf k}(t)\right>\,{\rm d}\tau,
\end{equation}
in which $\hat{\rho}_{\bf k}$ is the density operator
\begin{equation}
\hat{\rho}_{\bf k} = \sum_{i=1}^N e^{+i{\bf k}\cdot \hat{\bf r}_i}.
\end{equation}
This form of quantum mechanical intermediate scattering function has the same symmetry properties as the classical intermediate scattering function (both being real and even functions of $t$), it reduces to the classical intermediate scattering function in the high-temperature limit, and it is the intermediate scattering function that is directly approximated by methods like CMD and RPMD. (Indeed, the CMD approximation to $\tilde{F}(k,t)$ that we shall consider here is simply the classical intermediate scattering function on the centroid potential of mean force.)

Given this interpretation, Mountain's hydrodynamic result can be written as\cite{Mountain66} 
\begin{equation}
\int_0^{\infty} {\tilde{F}(k,t)\over \tilde{F}(k,0)} e^{-st}\,{\rm d}t = {n(k,s)\over d(k,s)},  
\end{equation}
where
\begin{equation}
n(k,s) = s^2+(a+b)k^2s+abk^4+c_{\rm s}^2(1-1/\gamma)k^2,
\end{equation}
and
\begin{equation}
d(k,s) = s^3+(a+b)k^2s^2+(c_{\rm s}^2k^2+abk^4)s+ac_{\rm s}^2k^4/\gamma.
\end{equation}
Here $a$, $b$, $c_s$ and $\gamma$ are the relevant physical parameters, which are related to various hydrodynamic and thermodynamic properties of the system: $a$ is the thermal diffusivity $a=\lambda/\rho c_{\rm V}$ that we have already mentioned, $b=(4\eta_{\rm s}/3+\eta_{\rm b})/\rho$ depends on the dynamic shear ($\eta_{\rm s}$) and bulk ($\eta_{\rm b}$) viscosities, $c_{\rm s}$ is the adiabatic speed of sound, and $\gamma = c_{\rm P}/c_{\rm V}$ is the ratio of the constant-pressure and constant-volume heat capacities.

Mountain himself,\cite{Mountain66} and most others since, have considered an approximate inversion of the Laplace transform in Eq.~(17) in which the roots of the cubic equation $d(k,s)=0$ are found to the lowest order in $k$. Schoen {\em et al.}\cite{Schoen86} went one step further and derived a correction to this leading order solution so as to satisfy the constraint that ${\rm d}\tilde{F}(k,t)/{\rm d}t=0$ at $t=0$. Both approaches are justified by the fact that Eq.~(17) itself is only valid in the limit as $k\to 0$, and they have the advantage of leading to convenient closed-form expressions for $\tilde{F}(k,t)$.\cite{Mountain66,Schoen86} However, we have found that when these expressions are used to fit simulation data to the hydrodynamic $\tilde{F}(k,t)$, the resulting $a$ and $b$ parameters have a $k$-dependence that is more pronounced than the inherent $k$-dependence that arises from the need to reach the hydrodynamic regime, which makes the extrapolation to infinite system size ($k=0$) more difficult than it need be. 

The way around this difficulty is to solve the cubic equation exactly, so as to obtain an expression for $\tilde{F}(k,t)$ that can be used as soon as $k$ enters the hydrodynamic regime. In physically relevant situations, the equation $d(s,k)=0$ will have one real root $s_0<0$, and a complex conjugate pair of roots $s_{\pm}=s_r\pm is_i$ with $s_r<0$.\cite{Hansen86} Assuming this to be the case, the inversion of the Laplace transform in Eq.~(17) gives
\begin{align}
{\tilde{F}(k,t)\over \tilde{F}(k,0)} &= {n(k,s_0)\over (s_0-s_+)(s_0-s_-)}e^{+s_0t} \nonumber \\
&+{n(k,s_+)\over (s_+-s_-)(s_+-s_0)}e^{+s_+t} \nonumber \\
&+{n(k,s_-)\over (s_--s_+)(s_--s_0)}e^{+s_-t} \quad \hbox{for}\ t\ge 0,
\end{align}
in which $s_0$, $s_+$ and $s_-$ are known functions of $ak^2$, $bk^2$, $c_{\rm s}k$, and $\gamma$. The first term in this expression is real and the remaining two are complex conjugates, so the overall expression is real, and one can show that it automatically satisfies the constraint that ${\rm d}\tilde{F}(k,t)/{\rm d}t=0$ at $t=0$. The expression is admittedly more complicated than the simple approximations given by Mountain\cite{Mountain66} and by Schoen {\em et al.},\cite{Schoen86} but it is no more difficult to use on a computer as a hydrodynamic model with which to fit simulation data to the parameters $a$, $b$, $c_{\rm s}$ and $\gamma$.

For this fitting, Cheng and Frenkel have made an important observation that we should now mention.\cite{Cheng20} Fitting data to a model with 4 parameters is never desirable, and here it is not necessary, because $c_{\rm s}$ and $\gamma$ are purely thermodynamic properties that can be determined independently. There are various ways to do this, both classically and quantum mechanically, such as using appropriate NVT estimators for quantities such as $c_{\rm V}$.\cite{Glaesemann02,Yamamoto05} However, the simplest way is probably just to use the thermodynamic relations\cite{Cheng20}
\begin{equation}
c_{\rm V} = {1\over N}\left({\partial E\over \partial T}\right)_V,
\end{equation}
\begin{equation}
c_{\rm P}-c_{\rm V}= -{T\over N}\left({\partial p\over\partial T}\right)_V^2\left({\partial p\over\partial V}\right)_T^{-1},
\end{equation}  
and
\begin{equation}
c_{\rm s}^2 = -{V^2\gamma\over Nm} \left({\partial p\over\partial V}\right)_T,
\end{equation}
to obtain $\gamma=c_{\rm P}/c_{\rm V}$ and $c_{\rm s}$ from the numerical partial derivatives of $E$ and $p$ with respect to the state variables $T$ and $V$. The classical estimators for $E$ and $p$ are well-known, and in a quantum mechanical PIMD calculation one can use the low-variance centroid virial energy and pressure estimators. 

The situation for CMD is slightly more subtle, because the centroid potential of mean force depends on the thermodynamic state point (see Sec.~II). It turns out that the correct way to proceed if one wants to calculate a $c_{\rm s}$ and a $\gamma$ that are consistent with the CMD calculation of $\tilde{F}(k,t)$ is {\em not} to change the centroid potential of mean force when calculating the numerical partial derivatives with respect to $T$ and $V$, just as one would not change the classical interaction potential when calculating these derivatives in the classical case. The resulting $c_{\rm V}$ from Eq.~(21) is what we shall refer to below as $c_{\rm V}({\rm CMD})$ -- the constant-volume heat capacity that is consistent with the dynamics obtained from CMD at the state point of interest.

\subsection{Results and Discussion}

In order to test this theory, we have used it to calculate the CMD approximation to the thermal diffusivity at each of the state points at which Roder and Diller measured the thermal conductivities of liquid para-hydrogen shown in Fig.~1. Since the results at all six state points tell the same story, it suffices to consider just one of them. For this, we have chosen the $T=22.001$ K, $\rho_{\rm m}=69.066$ kg m$^{-3}$ ($V=29.2$ cm$^{3}$ mol$^{-1}$) state point, at which the measured thermal conductivity is close to its maximum.\cite{Roder70} 

\begin{figure*}[t]
\centering
\resizebox{0.8\textwidth}{!} {\includegraphics{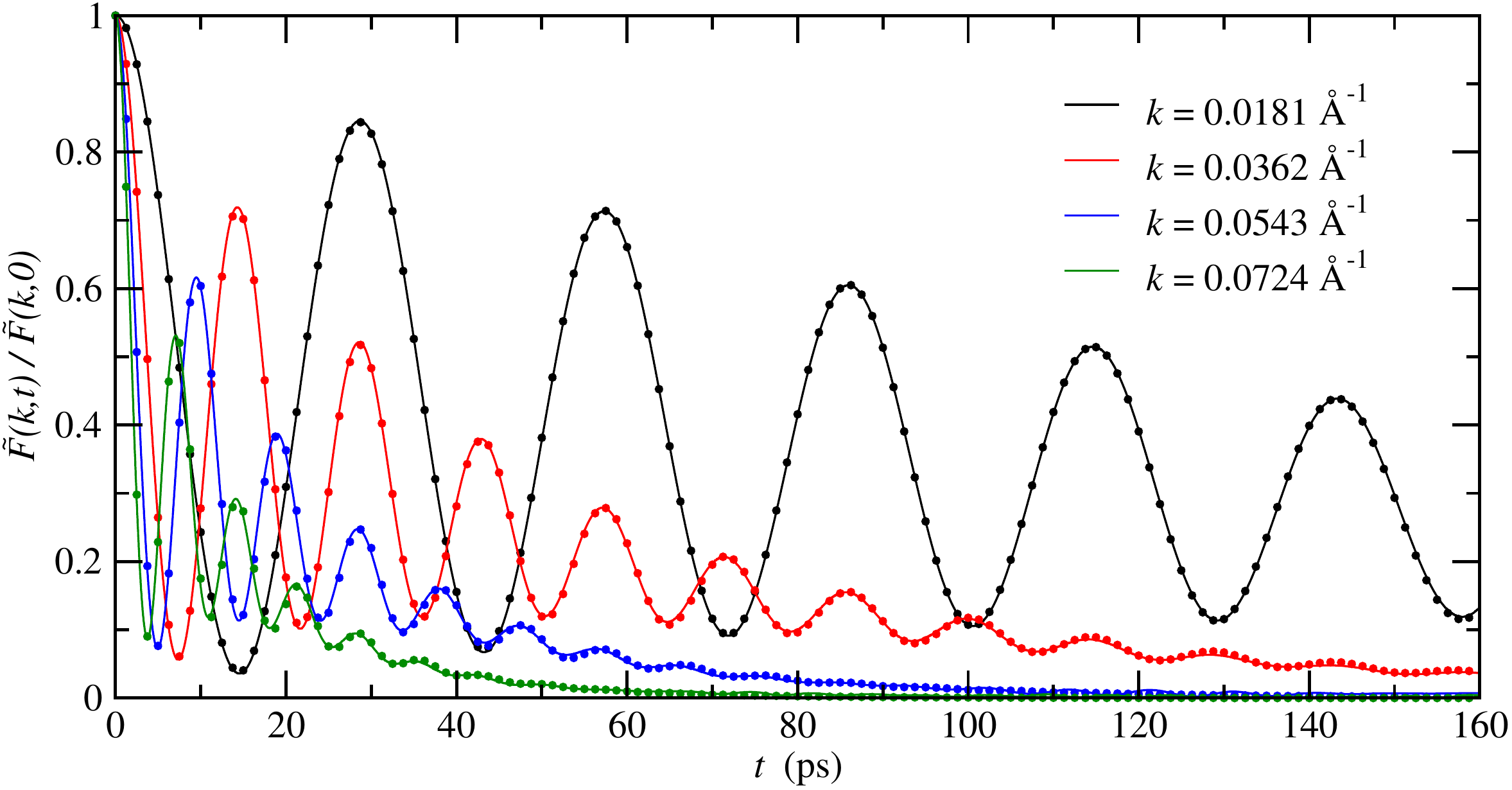}}
\caption{Normalised CMD intermediate scattering functions $\tilde{F}(k,t)/\tilde{F}(k,0)$ at the $T=22.001$ K, $\rho_{\rm m}=69.066$ kg m$^{-3}$ ($V=29.2$ cm$^{3}$ mol$^{-1}$) liquid para-hydrogen state point (thin solid lines), and fits to the hydrodynamic model in Eq.~(20) (dots).}
\label{SB_F_lambdas}
\end{figure*}

Fig.~5 shows the fits of four CMD intermediate scattering functions $\tilde{F}(k,t)$ to the hydrodynamic model in Eq.~(20) at this state point. The CMD calculations were performed by melting an initial face centred cubic lattice of para-hydrogen molecules containing $60\times 4\times 4$ unit cells in an initial NVT calculation on the centroid potential of mean force, and then running 3,000 independent NVE trajectories for a total of 320 ps with a resampling of the momenta between each one. The values of $c_{\rm s}$ and $\gamma$ used in the hydrodynamic fits were obtained from separate NVT simulations on a smaller ($4\times 4\times 4$) system, via the thermodynamic relations in Eqs.~(21) to (23). The temperature was controlled by applying a stochastic velocity rescaling thermostat\cite{Bussi07} at each temperature that was used to calculate the numerical partial derivatives of $E$ and $p$ with respect to $T$ and $V$. These calculations were found to give $c_{\rm V}({\rm CMD})=2.38\ k_{\rm B}$, $c_{\rm P}({\rm CMD})=4.83\ k_{\rm B}$, $c_{\rm s}({\rm CMD})=12.1$ \AA\ ps$^{-1}$, and $\gamma({\rm CMD}) = 2.03$. The corresponding path integral values obtained with $P=128$ ring polymer beads were found to be $c_{\rm V}({\rm PIMD})=1.44\ k_{\rm B}$, $c_{\rm P}({\rm PIMD})=2.40\ k_{\rm B}$, $c_{\rm s}({\rm PIMD})=10.8$ \AA\ ps$^{-1}$, and $\gamma({\rm PIMD}) = 1.67$. 

A comparison of these values shows that the classical thermodynamics on the (fixed) centroid potential of mean force does not agree especially well with the exact quantum thermodynamics, especially with regard to the two heat capacities. We believe that this is at the heart of the reason why the CMD Green-Kubo thermal conductivity results in Fig.~1 are in such poor agreement with experiment. The agreement of the CMD and PIMD heat capacities might be better if one were to allow for the temperature and density dependencies of the centroid potential of mean force when calculating $c_{\rm V}({\rm CMD})$ and $c_{\rm P}({\rm CMD})$, but then the resulting $c_{\rm s}$ and $\gamma$ would not be consistent with the CMD calculation of $\tilde{F}(k,t)$. The values of $c_{\rm s}$ obtained from the CMD and PIMD calculations differ by 12\%, and the values of $\gamma$ differ by 21\%. However, the values of the ratio $c_{\rm s}^2/\gamma$ differ by only 3\%. We believe that this is because this ratio is determined by the value of $\partial^2\tilde{F}(k,t)/\partial t^2$ at $t=0$ in the hydrodynamic model, when one takes the limit as $k\to 0$. Since this is a short-time dynamical property, it is likely to be captured more accurately than the other hydrodynamic parameters by CMD.

The fact that $c_{\rm s}({\rm CMD})$ and $\gamma({\rm CMD)}$ {\em are} consistent with the CMD calculation of $\tilde{F}(k,t)$ is clear from the quality of the hydrodynamic fits in Fig.~5. The remaining hydrodynamic parameters $a$ and $b$ were optimised separately for each value of $k$ considered in the figure. This was found to be necessary because there was still a residual $k$-dependence in $a$ and $b$ when Eq.~(20) was used for the hydrodynamic model. Presumably this indicates that our calculations were not yet fully in the hydrodynamic regime. The residual $k$-dependence in $a$ is illustrated in Fig.~6, which contains a linear least-squares extrapolation of $1/a$ to the limit of infinite system size ($k=0$). The error bars in this figure were obtained by dividing the 3,000 trajectories into 5 batches of 600, re-doing the hydrodynamic fit for each batch, and calculating the standard errors in the means of the resulting $1/a$ values at each $k$ point. The $k$-dependence of $a$ (or of the $1/a$ we have used, which gives a more compelling straight line extrapolation to $k=0$) is seen to be very slight -- less than 2\% over the whole range of $k$ values we have considered. This illustrates the advantage of using Eq.~(20) over an approximate inversion of the Laplace transform in Eq.~(17). [We initially used Schoen {\em et al.}'s expression\cite{Schoen86} for $\tilde{F}(k,t)/\tilde{F}(k,0)$ and found a significantly stronger $k$-dependence in the resulting parameters, which made the extrapolation to infinite system size more problematic. Cheng and Frenkel  experienced a similar issue in some of their calculations.\cite{Cheng20} Fig.~6 shows that Eq.~(20) solves this problem.]

\begin{figure}[b]
\centering
\resizebox{0.85\columnwidth}{!} {\includegraphics{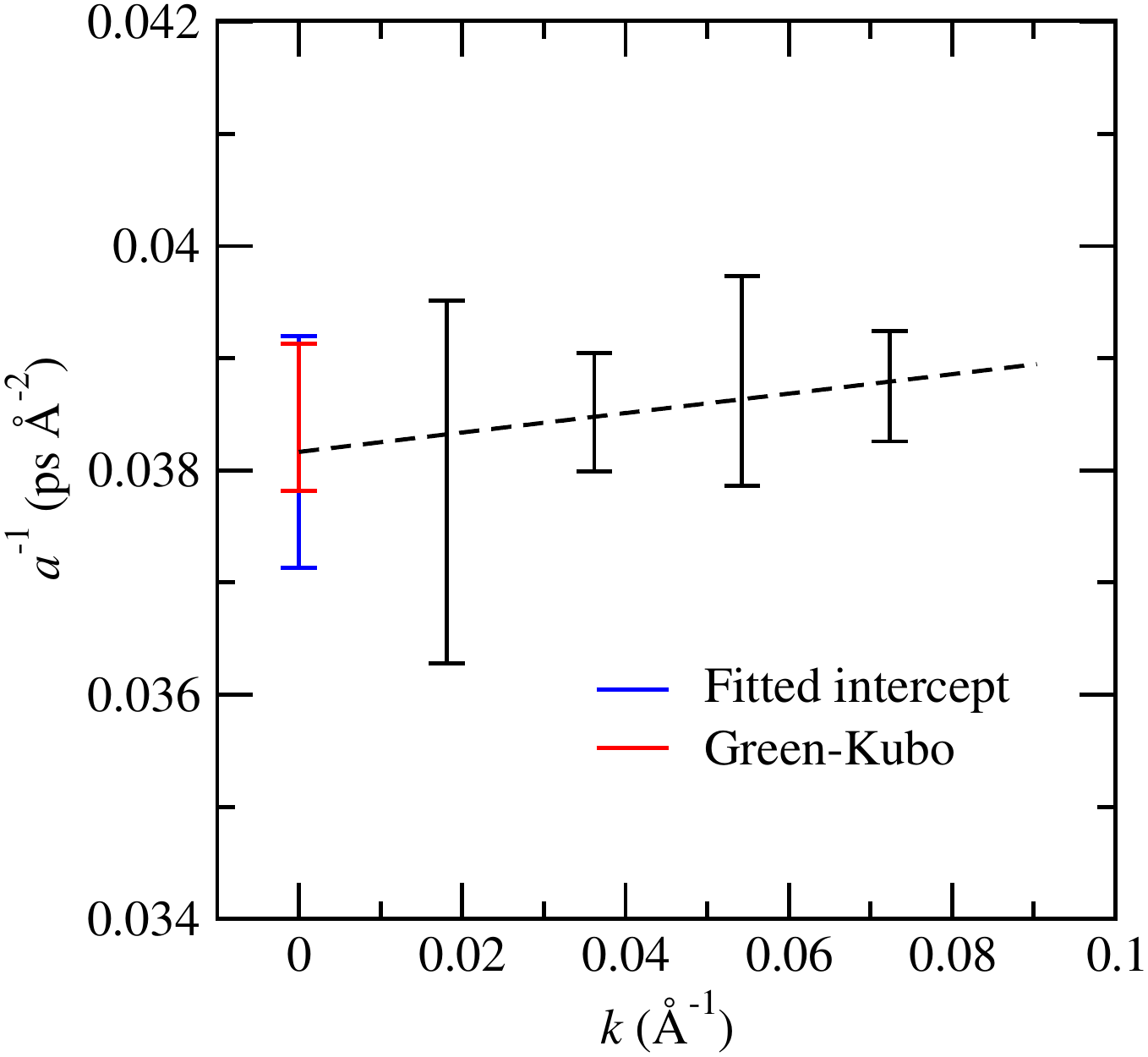}}
\caption{Extrapolation of the $1/a$ values obtained from the hydrodynamic fits in Fig.~5 to the limit of infinite system size ($k=0$). The red error bars at $k=0$ are those from the Green-Kubo method for $a({\rm CMD)}$ described in Sec.~IV. All error bars are one standard error in the mean.}
 \end{figure}

The final value of the thermal diffusivity obtained from the $k=0$ intercept in Fig.~6 is $a=26.2$ \AA$^2$ ps$^{-1}$. When this is combined with the density of the simulation ($\rho=0.02064$ \AA$^{-3}$) and the constant-volume heat capacity from the PIMD calculation ($c_{\rm V}=1.44$ $k_{\rm B}$), it gives a thermal conductivity $\lambda = \rho c_{\rm V}a$ of 0.107 W m$^{-1}$ K$^{-1}$, which is within 5\% of the experimental result of Roder and Diller\cite{Roder70} at this state point (0.1019 W m$^{-1}$ K$^{-1}$). However, if $a$ is combined with the constant-volume heat capacity of the CMD calculation ($c_{\rm V}=2.38\ k_{\rm B}$), the result is $\lambda = 0.177$ W m$^{-1}$ K$^{-1}$, which is more consistent with the CMD results of Yonetani and Kinugawa in Fig.~1. This clearly shows that the problem with the Green-Kubo CMD calculation of the thermal conductivity lies in the CMD heat capacity $c_{\rm V}$ and not the CMD thermal diffusivity $a$, which seems in this case to be rather accurate. In fact, a 5\% error in $a$ is surprisingly accurate given that the $c_{\rm s}$ and $\gamma$ parameters that were used to fit the CMD data to the hydrodynamic model in Fig.~5 disagree with those obtained from PIMD simulations by 12\% and 21\%. The CMD approximation is clearly more accurate for some of the parameters in the hydrodynamic model (recall the 3\% error in $c_{\rm s}^2/\gamma$ discussed above) than it is for others (e.g., the 21\% error in $\gamma$), but what really matters here is that it seems to be reasonably accurate for the thermal diffusivity.

The upshot of all this is that it is indeed possible to use CMD to calculate a thermal diffusivity $a$ from the decay of the equilibrium density fluctuations in the intermediate scattering function, and that when this is combined with an accurate PIMD calculation of $c_{\rm V}$ it gives a thermal conductivity $\lambda=a\rho c_{\rm V}$ that agrees reasonably well with experiment. However, this is an expensive calculation, even with the \lq\lq fast" CMD method we have used here. Reaching the hydrodynamic regime and then extrapolating to $k=0$ requires simulations with large system sizes, even when one uses a \lq\lq tube" geometry to mitigate the expense ($60\times 4\times 4$ face centred cubic unit cells still contain a total of 3,840 para-hydrogen molecules). The need to calculate $\tilde{F}(k,t)$ sufficiently accurately and over a sufficiently long time scale to give a credible fit to the hydrodynamic parameters also requires long simulations, such as the 0.96 $\mu$s simulations we have performed to construct Fig.~5. It would therefore be useful if there were a cheaper way to calculate $a({\rm CMD})$, and in the next section we shall show that there is (at least for systems like liquid para-hydrogen with simple pairwise interactions).

\section{Thermal diffusivity from the Green-Kubo relation}

\subsection{Theory}

One of the key things we have shown in Sec.~III is that the dynamics that gives the CMD intermediate scattering function $\tilde{F}(k,t)$ is consistent with the heat capacity we have called $c_{\rm V}({\rm CMD)}$ -- the heat capacity obtained by evaluating Eq.~(21) classically on the centroid potential of mean force {\em without} allowing for the temperature-dependence of the potential of mean force when evaluating the temperature derivative. This heat capacity is the same as the one that would be obtained from the standard classical expression
\begin{equation}
c_{\rm V} = {\left<(\Delta E)^2\right>\over Nk_{\rm B}T^2}
\end{equation}
in an NVT simulation on the centroid potential of mean force at the state point under investigation. That this $c_{\rm V}$ is consistent with the CMD calculation of $\tilde{F}(k,t)$ follows because it was an essential ingredient in the calculation of the $c_{\rm s}$ and $\gamma$ parameters that were used in the hydrodynamic fits in Fig.~5. If any other $c_{\rm V}$ had been used these fits would not have been so compelling.

An interesting implication is that $c_{\rm V}({\rm CMD})$ should therefore also be consistent with the dynamics in a CMD calculation of the thermal conductivity from the Green-Kubo relation, just as all aspects of a purely classical molecular dynamics simulation are internally consistent. Assuming this to be the case, we should be able to calculate the CMD thermal diffusivity from the Green-Kubo relation using $a({\rm CMD})=\lambda({\rm CMD)}/\rho\,c_{\rm V}({\rm CMD)}$, and thereby avoid the expense of calculating $a$ from the decay of equilibrium density fluctuations.

In the present \lq\lq fast" CMD context, in which CMD is simply classical molecular dynamics on the centroid potential of mean force in Eq.~(10), the Green-Kubo calculation of $\lambda$ is especially easy. One simply evaluates the classical expression for the thermal conductivity,
\begin{equation}
\lambda = {1\over 3Vk_{\rm B}T^2}\int_0^{\infty} \left<{\bf J}(0)\cdot{\bf J}(t)\right>\,{\rm d}t,  
\end{equation}
using the centroid potential of mean force, on which the energy current ${\bf J}$ is
\begin{equation}
{\bf J} = {1\over m}\sum_{i=1}^N (E_i+\boldsymbol{\sigma}_i)\,{\bf p}_i
\end{equation}
with
\begin{equation}
E_i = {|{\bf p}_i|^2\over 2m}+{1\over 2}\sum_{j\not=i}^N v^{(c)}(r_{ij}),
\end{equation}
and
\begin{equation}
\boldsymbol{\bf \sigma}_{i} = -{1\over 2}\sum_{j\not=i}^N {{\bf r}_{ij}{\bf r}_{ij}^T\over r_{ij}}{{\rm d}v^{(c)}(r_{ij})\over {\rm d}r_{ij}}.
\end{equation}
The CMD heat capacity $c_{\rm V}$ can then be calculated in a classical NVT simulation with the same centroid potential of mean force [using Eq.~(24)], and this can be combined with the CMD approximation to $\lambda$ to give $a({\rm CMD)}=\lambda({\rm CMD})/\rho c_{\rm V}({\rm CMD)}$.

\subsection{Results for liquid para-hydrogen}

\begin{figure}[t]
\centering
\resizebox{0.85\columnwidth}{!} {\includegraphics{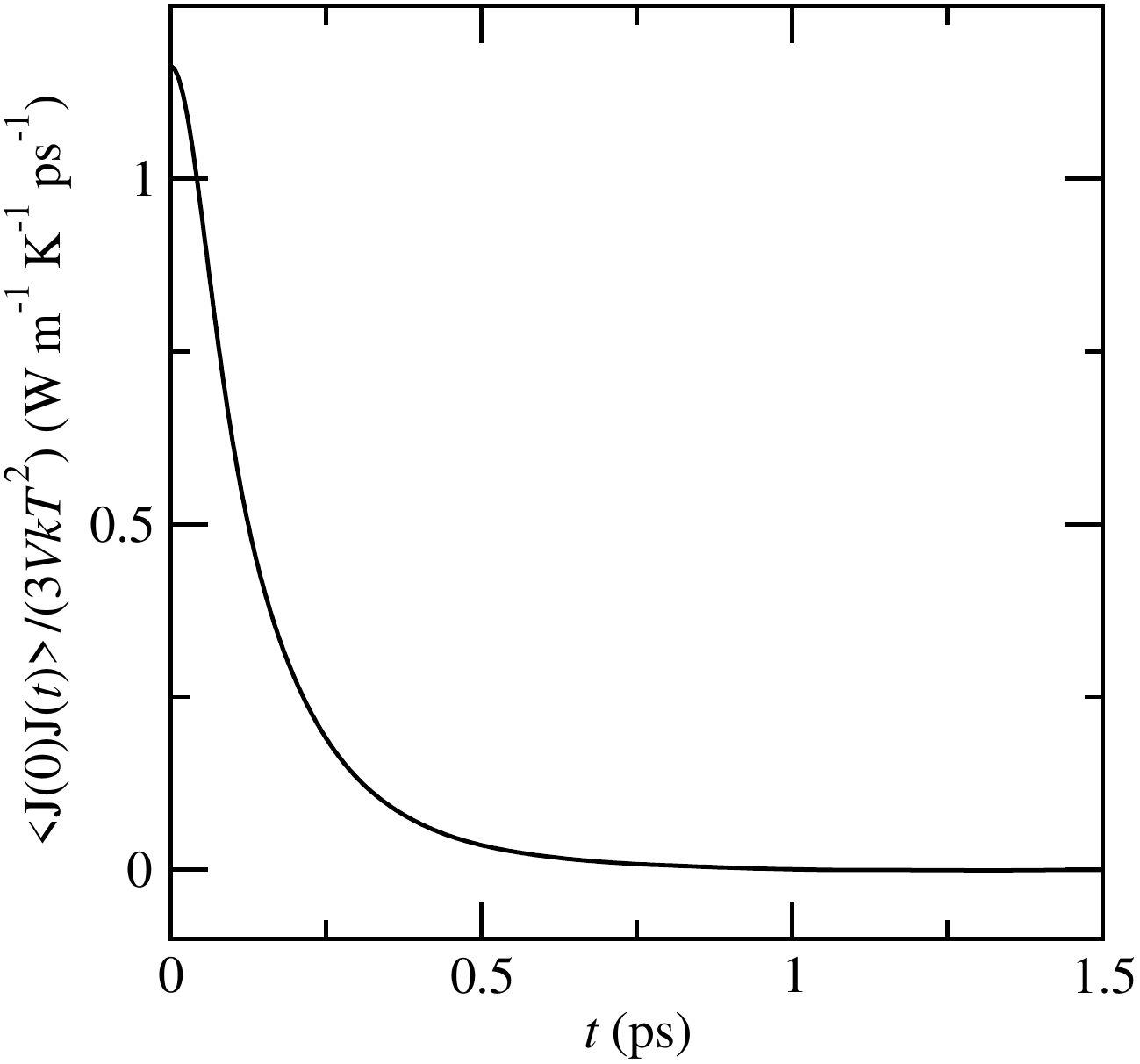}}
\caption{Scaled CMD energy current autocorrelation function at the $T=22.001$ K, $\rho_{\rm m}=69.066$ kg m$^{-3}$ ($V=29.2$ cm$^{3}$ mol$^{-1}$) liquid para-hydrogen state point.}
 \end{figure}

Fig.~7 shows the energy current autocorrelation function $\left<{\bf J}(0)\cdot{\bf J}(t)\right>$ obtained from a CMD simulation of liquid para-hydrogen at the $T=22.001$ K, $\rho_{\rm m}=69.066$ kg m$^{-3}$ state point considered in Sec.~III. This simulation was performed by melting a crystal of $6\times 6\times 6$ face centred unit cells in an initial NVT equilibration phase, and then running 50 independent 1.25 ns NVE trajectories on the centroid potential of mean force with a resampling of the momenta between each one. Because of the smaller system size and the shorter overall simulation time, these calculations were considerably less expensive than those used to calculate the CMD approximation to $\tilde{F}(k,t)$ in Sec.~III. Notice in particular that the correlation time of the energy current autocorrelation function in Fig.~7 is comparable to that of the velocity autocorrelation functions in Fig.~4, and orders of magnitude smaller than that of the intermediate scattering functions in Fig.~5. 

\def\s{\phantom{xx}}
\begin{table*}[!htbp]
\begin{center}
\caption{Summary of the present thermal conductivity, thermal diffusivity, and heat capacity results for liquid para-hydrogen, and comparison with the experimental measurements of Roder and Diller (Ref.~\onlinecite{Roder70}). (The numbers in parentheses denote the statistical errors in the final digits from our calculations.)}
\bigskip
\begin{tabular}{cccccccc} \hline
\s\s $T$\s\s & $\s\ \rho_{\rm m}$\s & $\s \lambda({\rm CMD})$\s & $c_{\rm V}({\rm CMD})$ &\s $a({\rm CMD})$\s & $c_{\rm V}({\rm PIMD}) $ &\s $\lambda({\rm Scaled})$\s&\s $\lambda({\rm Expt})$\s \\
K & kg\,m$^{-3}$ & Wm$^{-1}$K$^{-1}$ & $k_{\rm B}$ & \AA$^2$\,ps$^{-1}$ & $k_{\rm B}$ & Wm$^{-1}$K$^{-1}$ & Wm$^{-1}$K$^{-1}$ \\ \hline
16.942 & 74.556 & 0.209(3) & 2.703(6) & 25.1(4) & 1.307(3) & 0.101(2) & 0.0942 \\
19.522 & 71.799 & 0.190(3) & 2.522(4) & 25.4(4) & 1.404(3) & 0.106(2) & 0.0983 \\
22.001 & 69.066 & 0.176(3) & 2.377(4) & 26.0(4) & 1.439(4) & 0.106(2) & 0.1019 \\
24.990 & 64.916 & 0.151(3) & 2.228(5) & 25.3(5) & 1.495(7) & 0.101(2) & 0.1013 \\
30.026 & 56.361 & 0.119(2) & 2.068(5) & 24.8(4) & 1.535(3) & 0.088(2) & 0.0915 \\
33.001 & 45.900 & 0.091(1) & 2.004(6) & 23.9(4) & 1.563(4) & 0.071(1) & 0.0840 \\
\hline
\end{tabular}
\end{center}
\end{table*}

The thermal conductivity obtained from the area under the curve in Fig.~7 is $\lambda({\rm CMD}) = 0.176$ W m$^{-1}$ K$^{-1}$. When this is combined with the CMD heat capacity and the density of the simulation, the result is $a({\rm CMD)} = \lambda({\rm CMD})/\rho c_{\rm V}({\rm CMD)} = 26.0$ \AA$^2$ ps$^{-1}$. This is in excellent agreement with the CMD thermal diffusivity obtained from the $\tilde{F}(k,t)$ calculation in Sec.~III when the latter is extrapolated to the limit of infinite system size. [See also the comparison including error bars from batches of trajectories in Fig.~6, and note that there is no need to extrapolate the Green-Kubo result to the limit of infinite system size because the limit as $k\to 0$ has already been taken in the derivation of Eq.~(25).\cite{Hansen86}] As a result, the thermal conductivity calculated as $\lambda = \rho\,c_{\rm V}({\rm PIMD})a({\rm CMD})$ is again in excellent agreement with the experimental measurement of Roder and Diller\cite{Roder70}  at this thermodynamic state point.

\begin{figure}[t]
\centering
\resizebox{0.8\columnwidth}{!} {\includegraphics{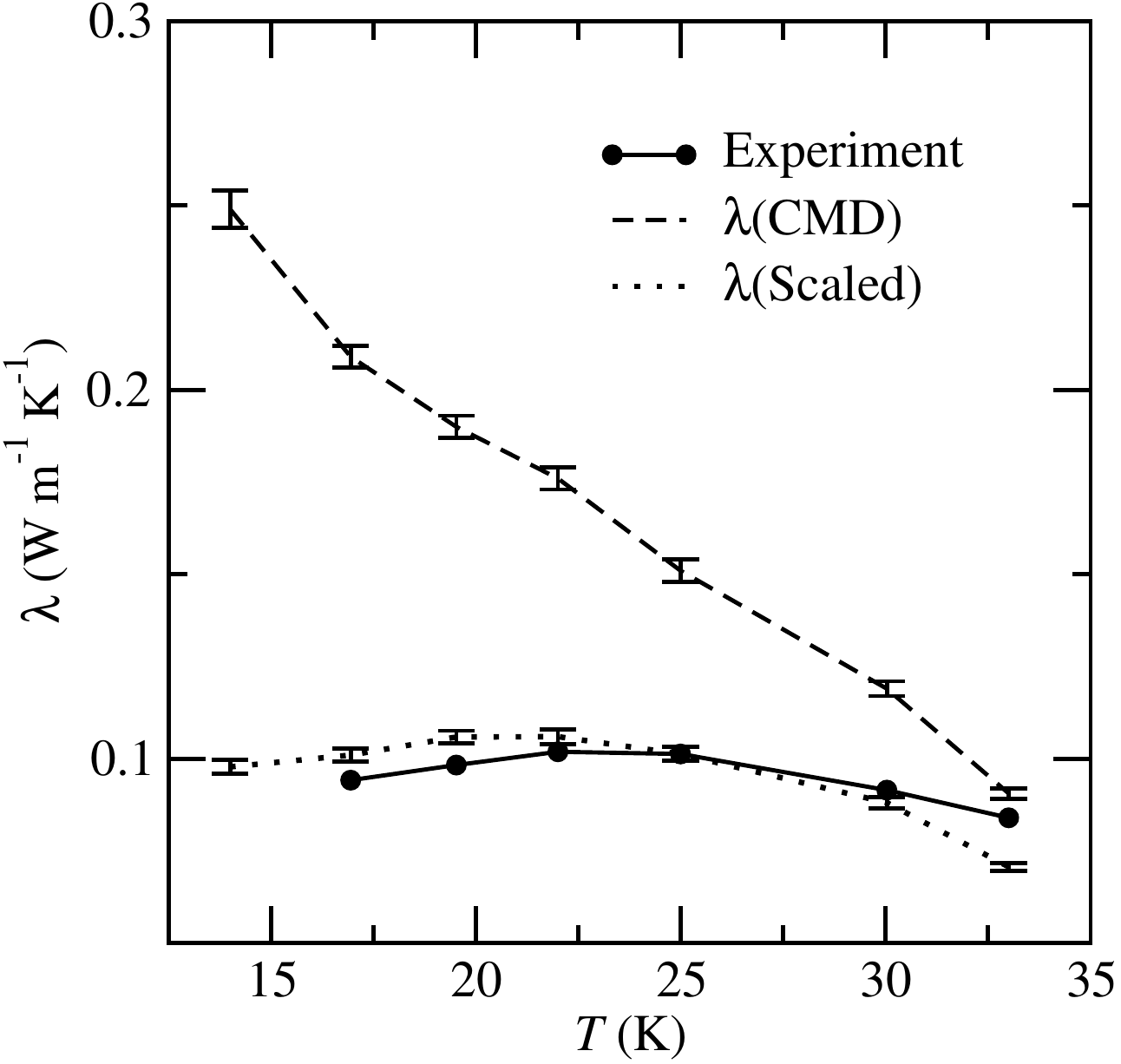}}
\caption{Comparison of the present CMD, $c_{\rm V}({\rm PIMD})/c_{\rm V}({\rm CMD})$ scaled CMD, and experimental (Ref.~\onlinecite{Roder70}) thermal conductivities of liquid para-hydrogen at densities close to the saturation line.}
 \end{figure}

We have repeated this calculation at all of the other state points on the experimental curve in Fig.~1. The resulting values of $\lambda({\rm CMD)}$, $c_{\rm V}({\rm CMD)}$, $a({\rm CMD})$, $c_{\rm V}({\rm PIMD})$, and $\lambda({\rm Scaled}) = \rho\, c_{\rm V}({\rm PIMD})a({\rm CMD)}$ are collected together for reference and compared with the experimental results of Roder and Diller\cite{Roder70} in Table~II. $\lambda({\rm CMD})$ and $\lambda({\rm Scaled})$ are also compared with the experimental measurements in Fig.~8, which contains additional theoretical results at the 14 K state point considered in Sec.~II.

It is clear from Fig.~8 that the $c_{\rm V}({\rm PIMD})/c_{\rm V}({\rm CMD})$ scaling in $\lambda({\rm Scaled})$ does an excellent job of correcting the raw $\lambda({\rm CMD})$ result and bringing it in better agreement with experiment. The correction works over a range of temperatures that extends from just beyond the triple point ($T_{\rm t}=13.8$ K) to just below the critical point ($T_{\rm c}=33.1$ K). In particular, it results in a decrease in the predicted thermal conductivity at temperatures below around 22 K, as is seen experimentally. It is clear from Table~II that the decrease comes from changes in both the quantum mechanical heat capacity $c_{\rm V}({\rm PIMD})$ and the CMD thermal diffusivity $a({\rm CMD})$, the latter of which also passes through a maximum at around 22 K. The agreement of $\lambda({\rm CMD)}$ with experiment begins to deteriorate as one approaches the critical point, but that is perhaps not surprising because one would not expect the large density fluctuations in the sub-critical liquid to be captured accurately using a Silvera-Goldman pair potential that has been truncated and shifted at $r_{\rm cut}=15$ bohr.

\subsection{Results for normal liquid helium}

Finally, to check that the good agreement between $\lambda({\rm Scaled})$ and experiment in Fig.~8 is not simply a coincidence, we have done some additional calculations on normal liquid helium-4 at densities close to the saturated vapour line. For these calculations, we looked at a relatively narrow temperature range between 3 and 4 K, so as to avoid approaching either the critical point ($T_{\rm c}=5.2$ K) or the lambda transition ($T_{\lambda}=2.2$ K). The PIMD simulations were done with $P=128$ ring polymer beads, and the CMD simulations by running 1000 independent 10 ps trajectories on the resulting centroid potential of mean force. The calculations used the HFD-B2 interatomic potential of Aziz {\em et al.},\cite{Aziz92} and were performed for a periodically replicated system of $N=256$ helium atoms.

The results of these calculations are summarised in Table~III and plotted in Fig.~9, where they are compared with the experimental results of Donnelly and Barenghi.\cite{Donnelly98} One sees that the $c_{\rm V}({\rm PIMD})/c_{\rm V}({\rm CMD)}$ scaling in $\lambda({\rm Scaled})$ again does very well in correcting the raw CMD thermal conductivity $\lambda({\rm CMD})$ and bringing it closer to experiment. In particular, it leads to a thermal conductivity that increases monotonically with temperature over the range considered, as is seen in the experiment. Our raw CMD results have the opposite temperature dependence, as do the earlier CMD results for this problem of Imaoka and Kinugawa.\cite{Imaoka17} The LSC-IVR calculations of Liu {\em et al.} also failed to predict the correct temperature dependence,\cite{Liu11} which insofar as we are aware the present calculations are the first to capture correctly.

\def\s{\phantom{xx}}
\begin{table*}[!htbp]
\begin{center}
\caption{Summary of the present thermal conductivity, thermal diffusivity, and heat capacity results for normal liquid helium, and comparison with the experimental measurements of Donnelly and Barenghi (Ref.~\onlinecite{Donnelly98}). (The numbers in parentheses denote the statistical errors in the final digits from our calculations.)}
\bigskip
\begin{tabular}{cccccccc} \hline
\s\s $T$\s\s & $\s\ \rho_{\rm m}$\s & $\s \lambda({\rm CMD})$\s & $c_{\rm V}({\rm CMD})$ &\s $a({\rm CMD})$\s & $c_{\rm V}({\rm PIMD}) $ &\s $\lambda({\rm Scaled})$\s&\s $\lambda({\rm Expt})$\s \\
K & kg\,m$^{-3}$ & Wm$^{-1}$K$^{-1}$ & $k_{\rm B}$ & \AA$^2$\,ps$^{-1}$ & $k_{\rm B}$ & Wm$^{-1}$K$^{-1}$ & Wm$^{-1}$K$^{-1}$ \\ \hline
3.00 & 141.23 & 0.0446(7) & 2.470(2) & 6.16(10) & 0.898(10) & 0.0162(3) & 0.01717 \\
3.30 & 138.34 & 0.0424(5) & 2.383(2) & 6.19(07) & 0.967(11) & 0.0172(3) & 0.01815 \\
3.65 & 134.17 & 0.0381(5) & 2.293(1) & 5.96(08) & 1.068(10) & 0.0177(3) & 0.01902 \\
4.00 & 128.97 & 0.0357(7) & 2.217(3) & 6.01(12) & 1.120(11) & 0.0180(4) & 0.01985 \\
\hline
\end{tabular}
\end{center}
\end{table*}

\begin{figure}[t]
\centering
\resizebox{0.8\columnwidth}{!} {\includegraphics{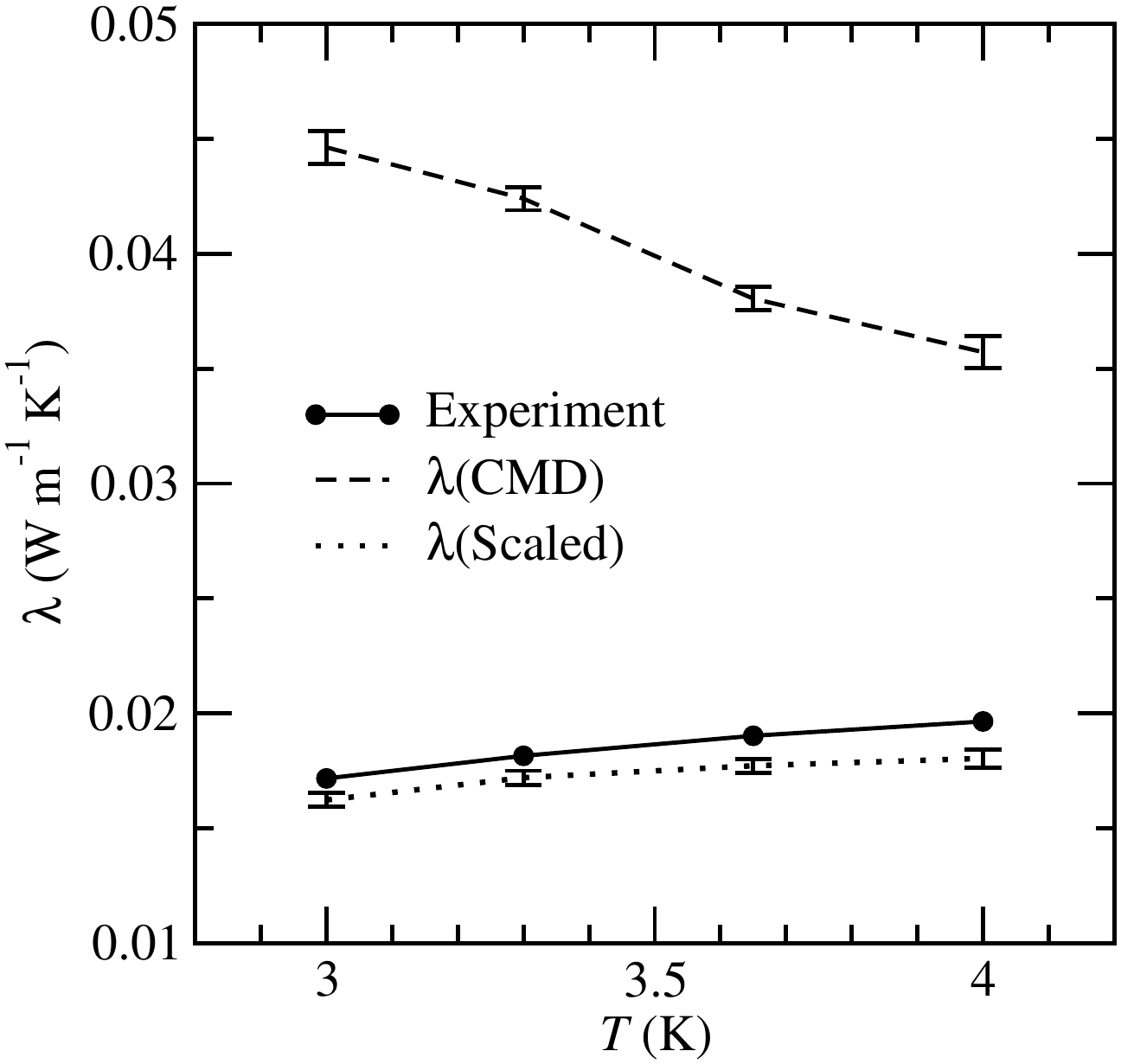}}
\caption{Comparison of the present CMD, $c_{\rm V}({\rm PIMD})/c_{\rm V}({\rm CMD})$ scaled CMD, and experimental (Ref.~\onlinecite{Donnelly98}) thermal conductivities of normal liquid helium at densities close to the saturation line.}
 \end{figure}

\section{Concluding Remarks}

While we have taken what might seem like a circuitous route to get there, via the discussion of equilibrium density fluctuations in Sec.~III [which we needed to establish that CMD provides a reliable way to calculate the thermal diffusivity $a$ and to establish precisely what we mean by $c_{\rm V}({\rm CMD})$], the final method we have suggested for the calculation of thermal conductivity in Sec.~IV is really very simple, and the results it has produced in Figs.~8 and~9 speak for themselves. These results are in better agreement with experiment than any other theoretical predictions we are aware of, both for liquid para-hydrogen and for liquid helium. We would therefore recommend this method for future calculations of thermal conductivity in situations where nuclear quantum effects play a significant role. 

The method boils down to calculating the CMD approximation to the Green-Kubo thermal conductivity $\lambda({\rm CMD)}$, and scaling the result by the ratio of the PIMD and CMD constant-volume heat capacities. For systems such as para-hydrogen and helium that can be described by pairwise interaction potentials, the calculation can be streamlined using the \lq\lq fast CMD" idea\cite{Hone05} described in Sec.~II. This makes it entirely practical and applicable to just about any system for which an ordinary classical molecular dynamics simulation would be feasible. For systems described by more complicated many-body interaction potentials, one could either try the more general force-matched implementation of fast CMD suggested by Hone {\em et al.},\cite{Hone05} or resort to a full-blown simulation with adiabatic CMD.\cite{Cao94d} This would be more expensive, but it does provide a practical way to calculate CMD thermal conductivities from the Green-Kubo relation, as Kinugawa and co-workers have shown in their simulations of para-hydrogen\cite{Yonetani04} and helium.\cite{Imaoka17} A cheaper alternative would be to develop the corresponding theory for the energy current in RPMD.\cite{Craig04} 

The Green-Kubo relation requires that the interaction potential can be decomposed into a sum of atomic contributions, which are needed to define the atomic energies $E_i$ that enter the expression for the energy current in Eq.~(26). It was thought for many years that this would not be possible if the forces on the atoms were calculated \lq\lq on the fly" using an {\em ab initio} method. However, this problem has recently been solved, both for solids in which the convective contribution to the energy current can be neglected,\cite{Carbongo17} and more generally.\cite{Marcolongo16} One might add that {\em ab initio} molecular dynamics simulations are now receding in popularity and being superceded by simulations on model potentials that have been machine learned from {\em ab initio} data.\cite{Behler07,Bartok10} This has the advantage that, because the machine-learned potential is cheaper to evaluate, one can run the simulation for more time steps and accumulate better statistics. In the present context, it has the added advantage that machine-learned potentials are almost always (if not always) written as a sum of atomic contributions, and are therefore ideally suited to the calculation of thermal conductivities from the Green-Kubo relation.\cite{Sosso12}

Given all of this, it is clear that there is a wide variety of potential applications of the method we have described in this paper. We have validated the method for quantum liquids, but there is no reason why it could not be applied to other disordered systems such as glasses. It could also be applied to crystalline solids at low temperatures, where nuclear quantum effects in the thermal conductivity are far from small\cite{Glassbrenner64} and the availability of anharmonic lattice dynamics simulations\cite{Ladd86,Turney09} would provide an interesting comparison. There are thus many possible avenues for further work.

\begin{acknowledgements} 
We are grateful to Mariana Rossi for her help in the early stages of this project and to Joseph Lawrence for his helpful suggestions throughout. Both made useful comments on the initial draft of this manuscript. Benjamin Sutherland is supported by the EPRSC Centre for Doctoral Training in Theory and Modelling in the Chemical Sciences, EPSRC Grant No. EP/L015722/1.
\end{acknowledgements}

\section*{Data Availability}

The data that support the findings of this study are available within the paper itself.

\end{document}